\documentstyle[12pt]{article}

\begin{document}
\thispagestyle{empty}
%
% ***** THE NEXT THREE LINES ARE FOR MADISON OUTPUT ONLY *****
%\font\fortssbx=cmssbx10 scaled \magstep2
%\hbox{ $\vcenter{\special{insert

%$disk1:[pheno.tex.inputs]uwlogo.imp}}}$
%\hskip.2in $\vcenter{\fortssbx University of
%Wisconsin - Madison}$ }
%
\hfill\vbox{\hbox{\bf MAD/PH/862}\hbox{January 1995}}
\vspace{.5cm}
\begin{center}
{\Large\bf NEUTRINOS FROM PRIMORDIAL BLACK HOLES}\\
\vspace{1cm}
{\large Francis Halzen and Bettina Keszthelyi}\\
{\it Physics Department, University of Wisconsin,
Madison, WI \ 53706}\\ \vspace{1.cm}
{\large Enrique Zas}\\
{\it Departamento de F\'\i sica de Part\'\i culas,
Universidad de Santiago}\\
{\it E-15706 Santiago de Compostela, Spain}%
\end{center}
\vspace{1.5cm}
\begin{abstract} \normalsize
The emission of particles from black holes created
in the early Universe has detectable astrophysical
consequences. The most stringent bound on their
abundance has been obtained from the absence of a
detectable diffuse flux of $100~$MeV photons. Further
scrutiny of these bounds is of interest as they, for
instance, rule out primordial black holes as a dark
matter candidate. We here point out  that these bounds
can, in principle, be improved by studying the diffuse
cosmic neutrino flux. Measurements of near-vertical
atmospheric neutrino fluxes in a region of low
geomagnetic latitude can provide a  competitive  bound.
The most favorable energy to detect a possible diffuse
flux of primordial black hole origin is found to be a
few MeV. We also show that measurements of the diffuse
$\nu _\tau$ flux is the most promising to improve the
existing bounds deduced from gamma-ray measurements.
Neutrinos from individual black hole explosions can be
detected in the GeV-TeV energy region. We find that the
kilometer-scale detectors, recently proposed, are able
to establish competitive bounds.
\end{abstract}

\section{Introduction}

Soon after Hawking described the quantum evaporation of
black holes~\cite{haw}, it was suggested that emission
from black holes, created in the early Universe, has
detectable astrophysical
consequences~\cite{pagehawk,carr75,carrlims}.
Such black holes are referred to
as primordial. They may be formed by the collapse of density
fluctuations at an early epoch~\cite{form1}, at a cosmic
phase transition~\cite{form2} or by more exotic mechanisms
such as the collapse of closed cosmic strings~\cite{form3}.
Attempts were made to search the Hawking
radiation from primordial black holes (PBHs). The instantaneous
emission spectra~\cite{haw} of particles by an  individual black
hole resemble a black body spectrum with temperature  inversely
proportional to the  black hole mass. Only holes of masses
below the critical mass $M_*\simeq 4-6 \times 10^{14}$~g
evaporate within the age of the Universe through this
mechanism.

In this paper we assume that  a particle species is
emitted when the black hole temperature exceeds the
relevant rest mass and if the particle appears elementary
at this temperature.
A solar mass black hole emits only massless particles,
photons and possibly neutrinos. As a black hole
evaporates its temperature rises crossing the emission
threshold of heavier particles so that new degrees of freedom
are added to the emission.  A black hole of mass $M_*$ also
emits electrons and positrons with energies in the $50-100$~MeV
energy range. This average energy is sufficiently close to
the muon and pion rest mass so that they are also kinematically
accessible. In the Standard Model this means that the quark and
gluon degrees of freedom will contribute to the radiation. When
the temperature approaches $\Lambda_{QCD}$ quarks and gluons
are emitted freely which subsequently fragment into hadrons by
mechanisms familiar from accelerator studies. Fragmentation
products decay into stable particles which add to the direct
Hawking emission, thus contributing an enhancement in
the spectrum near $\simeq 140$~MeV, the pion rest
mass~\cite{jane}.

PBHs can be searched for as nearby point sources or,
alternatively, one may search for the accumulated diffuse
flux of all PBHs which evaporated within the lifetime of
our Universe. Page and Hawking obtained the first such
bound from the requirement that this diffuse flux cannot
exceed the observed gamma ray spectrum~\cite{pagehawk}.
The most stringent bound is obtained for photons with
energy of order $100~$MeV. They only considered the
direct photon flux. This result was later updated to
include all Standard Model degrees of freedom as well
as the fragmentation of quarks and gluons~\cite{jane,halzas}.
Bounds obtained from the observed diffuse electron,
positron and antiproton spectrum~\cite{carrlims} are
remarkably close to those from gamma rays.

Searches for individual, nearby holes provide bounds on
explosion rates which are complementary to the diffuse
bounds. These bounds depend on the possible clustering
of PBHs around galaxies and on their final stage evolution
which is dictated by yet unsettled high energy physics.
Though the solutions to these problems are somewhat ambiguous,
an observation of such an explosion would reveal all the
degrees of freedom of all particles in the
Universe irrespective of their mass.

The extension of the work on photons to include neutrinos is most
natural. The construction of a new generation of solar and
supernova neutrino detectors as well as the commissioning of a
new type of high energy neutrino telescope provide us with novel
ways to search for PBHs. The low cross sections for
neutrino interactions require large detector volumes, but they
also allow detection on the Earth's surface avoiding shielding by
either the interstellar medium, the atmosphere or the Earth itself.
The emission of direct neutrinos has been previously addressed by
Carr~\cite{carrlims}. In this work we include the production of
neutrinos via the decay of hadrons. By comparing the diffuse
neutrino fluxes from PBHs to the observed neutrino
backgrounds we determine the most favorable energy regions for
obtaining new independent bounds on PBH abundances. We will show
that such regions of opportunity exist in the range of sensitivity
of both low- and high energy neutrino telescopes.

\section{Hawking radiation from PBHs and emission models}

Hawking showed that a black hole emits particles thermally with a
temperature that depends only on its mass, charge and angular
momentum. We can safely assume that PBHs, formed in the early
Universe, are neutral and nonrotating because any nonvanishing
initial charge or angular momentum would have been lost soon after
their formation. A black hole emits  particles with energy
in the range $(E, E+dE)$ at a rate \cite{haw}
\begin{equation}
{{\rm d^2}N\over {\rm d}t{\rm d}E} = {1\over 2\pi \hbar}~
{\Gamma_s (E,M) \over \exp \left(8\pi G M E/ \hbar c^3 \right)-
(-1)^{2s}}    \label{eq:haw}
\end{equation}
per particle degree of freedom. Here $M$ is the mass of the
hole, $s$ is the particle spin and $\Gamma _s (E,M)$ is the
absorption coefficient.

Although Eq.~(\ref{eq:haw}) does not represent a perfect
black body spectrum because $\Gamma _s(E,M)$ is energy
dependent, one can define a temperature
\begin{equation}
kT={\hbar c^3 \over 8\pi G M}.      \label{eq:temp}
\end{equation}
The absorption coefficient $\Gamma _s(E,M)$ depends on
$E$, $M$ and $s$ and any other internal degrees of freedom
as well as on the rest mass of the emitted particle. For
relativistic or massless particles $\Gamma _s (E,M)$ is
a function of $ME$; its exact form has to be calculated
numerically~\cite{page76}. It can be adequately approximated
in terms of the absorption cross section
$\sigma_s (E,M)$~\cite{jane}:
\begin{equation}
\Gamma _s(E,M) ={E^2 \sigma _s (E,M)\over \pi \hbar ^2 c^2},
\end{equation}
where $\sigma _s (E,M)$ can be replaced by its averaged value
\cite{jane}
\begin{equation}
{\bar \sigma }_s= {\int {\rm d} E~\sigma _s (E, M) \over \int{\rm d}E }
=\left\{ \begin{array}{ll}
        56.7\ {G^2M^2/ c^4} & s={1\over 2} \\
        20.4\  {G^2M^2/ c^4} & s=1        \\
        2.33 \ {G^2M^2/ c^4} & s=2 \ .
	\end{array}
	\right.
\end{equation}

Particle physics determines the number of states the black hole
emits  at a given temperature.  They
can be emitted directly by the Hawking mechanism, or by the decay and
fragmentation products of such particles. The first flux we will
refer to as ``direct", the
latter as ``indirect" fluxes.
As a consequence of emission the black hole loses mass at a rate
\begin{equation}
{{\rm d}M \over {\rm d} t}=-{\alpha (M) \over M^2},
\label{eq:massloss}
\end{equation}
where $\alpha (M)$ counts the degrees of freedom of the
emitted particles. As the black hole radiates, its
temperature rises at an increasing rate because
$\alpha (M) $ increases smoothly at the rest mass threshold
for each new massive particle~\cite{jane}. For the Standard
Model $\alpha (M)$ is shown in Fig.~1 of Ref.~\cite{halzas}.
The details of the final stage of the PBH evaporation thus
depend on the high energy value of $\alpha(M)$ which is
determined by the underlying particle physics. Whatever the
model, the black hole loses mass approximately as $M^{-3}$,
and its temperature increases accordingly. Its luminosity
increases roughly as $M^{-2}$ as $M$ decreases. We will refer
to this runaway process as explosion.

To a good approximation the solution of
Eq.~(\ref{eq:massloss}) for the evolution with time of the
mass $M(t)$ of a black hole with initial mass $M_i$ is
\begin{equation}
M(t)\simeq (M_i^3-3\alpha t )^{1/3}=M_*
\left( \left({M_i\over M_*}\right) ^3-{t\over t_0}\right)^{1/3} ,
\label{eq:mev}
\end{equation}
where $\alpha =\alpha (M_*)$ and $t_0$ is the age of the Universe.
We recall that $M_*$ is the mass of a black hole, formed
in the early Universe and evaporating today.
One can use Eq.~(\ref{eq:mev}) to relate the initial
mass $M_i$ of the black hole to the lifetime $\tau$ over
which it completely evaporates.

The most conservative choice for the underlying particle physics
model
is the Standard Model. It provides a lower bound on the high
energy value of $\alpha (M)$, not only because its degrees of
freedom are restricted to known particles with the exception
of the Higgs, but also because above the deconfinement
temperature, $T_{\scriptscriptstyle D} \sim \Lambda_{QCD}
\simeq 100-300~GeV$, quarks and gluons are emitted as fundamental
particles instead of pions and heavier hadrons. Quarks and
gluons subsequently fragment into (mostly) pions. Only pions
can be emitted directly below the deconfinement temperature,
assuming that their rest mass is somewhat smaller than
$\Lambda _{QCD}$. Other models may include more degrees of
freedom. For example supersymmetry increases $\alpha (M)$
by at least a factor of~3. Instead of trying to list all
the possibilities, we only mention the other extreme example,
the Hagedorn picture, where the number of degrees of freedom
increases exponentially. In Hagedorn-type models spectacular
explosions are obtained because of these exponentially
increasing hadron degrees of freedom. We consider these models
unrealistic in the framework of QCD and will not consider them
further. Throughout this paper we will assume the degrees of
freedom to be those of the Standard Model, even at the highest
temperatures, and in our calculations we will use
$T_{\scriptscriptstyle D}=200$~GeV, $m_{\rm top}=170$~GeV and
$m_{\rm Higgs}=100$~GeV.

\section{Neutrino fluxes}

We are now ready to calculate the neutrino fluxes for each
species. The direct neutrino flux, which is the same for all
neutrino  flavors, is given by Eq.~(\ref{eq:haw}).
The indirect neutrino fluxes are flavor dependent and fall into
two categories. They can be the decay products of directly
emitted heavier particles, like muons or taus.
Above the QCD deconfinement temperature they can  also be
produced by the decay of quark and gluon fragmentation products,
mostly pions in quark jets. These fluxes are studied in  detail
in \cite{jane}. Here we shall make the adequate approximation
that all the  fragmentation products are pions,
hence the decay of fragmentation products only contributes to
the muon  and electron neutrino flux through charged pion decay.

Our main results are independent of the explicit choice of
the quark fragmentation function. We use the empirical
fragmentation function~\cite{schramm}
\begin{equation}
{{\rm d} N_{\pi}\over {\rm d}z}={15\over 16}(z-1)^2 z^{-3/2},
				\label{eq:frag}
\end{equation}
where $z=E_{\pi}/E$ and $E$ is the quark (jet) energy.
This choice of the fragmentation function is convenient
because it yields realistic, analytic expressions for the
hadron decay fluxes when we further approximate the flux by
a delta function at its peak value \cite{halzas}.

Two thirds of the pions are charged and decay into
$\bar{\nu}_{\mu}+\mu$ and their charge conjugates.
The direct quark emission is convoluted with the fragmentation
function and the pion decay distributions. Naturally, whenever
fragmentation is considered, the flux depends on the PBH mass
at the moment of emission. The muons from pion decay also
decay ($\mu^{-} \rightarrow \nu_{\mu}~\bar{\nu}_e~e^{-}$),
and thus contribute to the $\nu_{\mu}$ and the $\nu_e$
fluxes. To calculate the contributions of such decays
the convolution must therefore also involve muon decay.

The decay distribution of a particle from any decay is given by:
\begin{equation}
{{\rm d}n \over {\rm d}E} ={1\over \Gamma }{{\rm d}\Gamma
\over {\rm d} E}~,
\end{equation}
where $\Gamma $ is the decay width. The distribution is
normalized to the number of identical particles produced by
a single decay. The calculations are straightforward. Below
we explicitly list the distributions used because they
correct errors in an earlier calculation~\cite{jane}:

Pion decay: $\pi^{\pm}~\rightarrow ~\mu ^{\pm} ~\nu_\mu
(\bar{\nu}_\mu)$.
The $\nu_\mu$ distribution  is given by:
\begin{equation}
{{\rm} dn_{\nu_\mu}\over {\rm d} E_\nu}=\delta
\left(E_\nu -  ({m_\pi^2-m_\mu^2
\over 2 m_\pi} ) \right),
\end{equation}
in the rest frame of the pion ($\beta =0$), and
\begin{equation}
{{\rm} dn_{\nu_\mu}\over {\rm d} E_\nu}=
{m_\pi^2\over m_\pi^2  -m_\mu^2 }
{1\over \sqrt{E_\pi^2 -m_\pi^2}}\ ,
\end{equation}
where the allowed range of $E_\nu$ is
\begin{equation}
{{1\over 2}~( m_\pi ^2 - m_\mu ^2 )
\over E_\pi +\sqrt{E_\pi^2 -m_\pi^2}}
\leq E_\nu \leq
{{1\over2}~( m_\pi ^2 - m_\mu ^2 )
\over E_\pi -\sqrt{E_\pi^2 -m_\pi^2}}
\end{equation}
in the lab frame ($\beta \neq 0$).

Muon decay: $\mu^{\pm} \rightarrow {\bar \nu}_\mu(\nu_{\mu})
{}~\nu_e(\bar{\nu}_e)~e^{\pm}$ The distribution of the muon
neutrinos is given by:
\begin{eqnarray}
{{\rm d} n_{\nu_\mu}\over {\rm d}E_\nu}
= {2\over \gamma m_\mu } \times \left\{ \begin{array} {ll}
{1\over \beta }   \left[{5\over 6} -{3\over 2}\epsilon^2+
{2\over 3} \epsilon^3 \right]&  \quad   {\rm for}   \
	{1-\beta \over 1+\beta } \leq \epsilon \leq 1
\ , \\                            &              \\
	{2\left( 1+\beta \right) ^2 } \left[ 3\epsilon^2-
{2\over 3} \left({3 +\beta ^2 \over 1 - \beta}\right)
\epsilon ^3 \right]&    \quad {\rm for}  \
	 0 \leq \epsilon \leq {1-\beta \over 1+\beta } \ .
	\end{array}
	\right.
\end{eqnarray}
Similarly the electron neutrino distribution is:
\begin{eqnarray}
{{\rm d} n_{\nu_e}\over {\rm d}E_\nu} = {2\over\gamma m_\mu}
\times \left\{ \begin{array} {ll}
{1 \over \beta } \left[ 1 -3 \epsilon ^2
+ 2 \epsilon^3 \right]&  \quad  {\rm for} \
	{1-\beta \over 1+ \beta } \leq \epsilon \leq 1
\ ,    \\                         &              \\
	{4 \left( 1+\beta \right) ^2 } \left[3
\epsilon ^2 - \left({3+\beta ^2 \over 1 - \beta} \right)
\epsilon ^3 \right]&    \quad      {\rm for} \
	0 \leq \epsilon \leq {1-\beta \over 1+\beta }\ .
	\end{array}
	\right.
\end{eqnarray}
In both the above cases $\epsilon=E_\nu / {E_\nu^{\rm max}}$~,
where
\begin{equation}
E_{\nu}^{\rm max}={\gamma m_\mu\over 2} (1+\beta),
\end{equation}
is the maximum neutrino energy,
and the electron mass is neglected.

Tau neutrinos represent a somewhat different story. They can
hardly be produced in hadron decays since decays of hadrons
into ${\nu}_{\tau}$ are highly suppressed. Decays into
$\nu_{\tau}~\bar{\nu}_{\tau}$ proceed  via $Z^0$ exchange and
therefore are suppressed with respect to decays mediated by
a photon.  Decays into a $\tau-\nu_{\tau}$
or $\tau-\bar{\tau}$ pairs are kinematically forbidden except
for heavy hadrons which rarely appear as quark or gluon
fragmentation products. Moreover, the decays into
$\tau-\nu_{\tau}$ channels have small branching ratios.
In the end most of the ${\nu}_{\tau}$ flux is direct. The
only other significant contribution comes from the decay
of direct  $\tau$'s.

For the neutrino fluxes from $\tau$ decay we estimate that
$\sim$~40\% of the decays occur in leptonic modes, with the
rest in semi-hadronic modes. In the free quark approximation
this comes from the relative number of colors (3) to
neutrino flavors (2) into which the decay may proceed.
The number distributions for each type of particle  are
the same as the above three body  distribution in the limit
of massless final states. Thus an approximate  $\nu _\tau$
flux is given by replacing $m_\mu$ by $m_\tau$ in the above
formulas. The $\nu _e$ and $\nu _\mu $ fluxes coming from
$\tau \rightarrow e {\bar \nu}_e \nu_\tau$ and $\tau
\rightarrow  \mu {\bar \nu}_\mu \nu_\tau$, respectively,
are  each  20\% of the total
$\nu _\tau$ flux from $\tau $ decay.

In the following sections we will compute $f(x, M)$, the
total neutrino flux of energy
$E=\left(8\pi GM/\hbar c^3 \right)^{-1}x$ for a given
species. For $\nu_\mu$, the direct, the muon decay, and
the fragmentation flux contribute to $f(x, M)$. The direct
flux depends on $x$ only, while the fragmentation flux
depends also on the PBH mass $M$ at the time of the emission.

\section{Diffuse neutrino fluxes}

Our calculation of the diffuse neutrino spectrum closely
parallels  that of Ref.~\cite{pagehawk} for gamma rays. In
order to calculate the present neutrino flux of energy $E_0$
we must integrate the contributions over cosmological time
$t$ for all PBH of mass $M$ at the blue shifted energy $E$
\begin{equation}
E = {E_0 \over r},
\end{equation}
where $r\equiv R/R_0=(1+Z)^{-1}$, $Z$ is the redshift at
time $t$. The subscript $0$ denotes the value of a quantity
at the present epoch. We assume a standard Friedman
cosmological model. Thus the Universe is radiation dominated
for $t< 10^{-6}~s$ and matter dominated for $t>10^{-6}~s$.

Assuming a uniform distribution of PBHs, created shortly
after $t=0$, the initial number of PBHs per comoving volume
of mass less than $M_i$ is
\begin{equation}
n(M_i)\equiv {\cal N} \int _0^{M_i/M_*} {\rm d}y~s(y)~,
\end{equation}
where $y=M_i/M_*$ and $s(y)$ is a dimensionless function
that determines the form of the initial mass distribution.
Here $s(1)=1$. ${\cal N}$ is the initial number density of
PBH per logarithmic mass interval at $M=M_*$. From various
cosmological considerations we know that
${\cal N} \leq 10^4~pc^{-3}$. PBH clustering in the galaxy
can increase their local density by as much as a factor of
$\xi =10^7$ which dramatically improves our chances to
observe them~\cite{halzas}. In our calculations we will
not assume any clustering, {\it ie.} $\xi =1$.

If PBHs form from scale invariant density perturbations
with a Gaussian spectrum, then
\begin{equation}
s(y)=y^{-\beta}, \label{eq:powlow}
\end{equation}
where $2<\beta<3$, depending on the equation of state of
the Universe at the formation of PBHs \cite{carr75}. If
$p=\gamma \rho $, where $p$ is the presure and $\rho $
is the energy density, then
\begin{equation}
\beta = {1+3\gamma \over 1+\gamma }+1 \ .
\end{equation}

Let $f(x,M)$ represent the total neutrino flux of energy
$E_\nu$ as defined in the previous section, then the diffuse
flux per unit area today is:
\begin{equation}
{d^4N_{\nu} \over dA~d\Omega~dt~dE_{\nu 0}}
 ={{\cal N}c\over 4\pi} \int_0^{t_0} {\rm d} t' (1+Z)\int
{\rm d} y~s(y)~f(x,M) \ , \label{eq:diff0}
\end{equation}
where $y$ is to be integrated over all values of the initial
masses of the PBHs present at $t'$, and $x$ is the value
of $\left( 8\pi G M/\hbar c^3\right)E_\nu$ at $t'$ and $y$.
Absorption on ambient  matter in the Universe should be taken
into account. For neutrinos
this is negligible. Making use of (\ref{eq:mev}) and
(\ref{eq:powlow}), the integral (\ref{eq:diff0}) becomes:
\begin{eqnarray}
{d^4N_{\nu} \over dA~d\Omega~dt~dE_{\nu 0}}
&=&{{\cal N}c \over 4 \pi x_*^3} \int_0^1 dt'~r^2
\int_0^{\infty} dx~x^2   \nonumber \\
& &                             \\   \label{eq:diff}
& &\qquad \times f(x,M)\left({t' \over t_0}
+ {r^3 x^3 \over x_*^3}
\right)^{-({\beta+2 \over 3})}, \nonumber
\end{eqnarray}
where $x_*=\left(8\pi G M_*/\hbar c^3\right) E_\nu$.
The right hand side can be integrated numerically once the
contributions to
$f(x,M)$ are specified. For electron neutrinos we include
the contribution from the decay of secondary muons (from
the decay of fragmentation pions), the decay of direct
muons and taus and the direct flux. For muon neutrinos we
include the contributions from the decay of fragmentation
pions, the decay  of secondary muons, the decay of direct
muons and taus and the direct flux. In the fragmentation
case the $x$-integration limits are modified. For tau
neutrinos only the direct flux and the flux from direct
tau decay contribute significantly as discussed earlier.

In Figs.~1 and 2 we plot the resulting fluxes. The fluxes
have been  calculated for a PBH density taken to be the
maximum allowed by bounds derived from gamma ray measurements
\cite{halzas}. The main characteristics of the neutrino
fluxes are: i) in the low energy  region the slopes of the
curves are $\beta$ dependent, while at high energies they
decrease as $E^{-3}$, independent of $\beta $, ii) all fluxes
display a ``knee'' (a change of slope) in the 50-100~MeV
region corresponding to the peak emission energy of black
holes of mass  $M_*$, iii) the fragmentation spectra display
a knee at around  10~MeV and, finally, iv) the direct spectra
are negligible above the knee and dominate in the region below.

\section{Background fluxes and bounds on PBH abundances}

It is in principle possible to obtain an improved bound
whenever the predicted neutrino fluxes, computed for the
maximal PBH abundance allowed by the gamma-ray bound, exceed
the neutrino backgrounds.  These are predominantly of solar
and atmospheric origin and can be separated by energy region.

The solar flux extends to 18.8~MeV where it is sharply cut off.
Below the cutoff it dominates with a very featured spectrum
which traces the  different nuclear reactions in the Sun's
interior. The reaction yielding the lowest flux but the most
energetic neutrinos is $He + p \rightarrow \nu_e +\ ...$,
usually abbreviated $hep$. This flux is large compared to the
diffuse PBH flux but it has two distinguishing characteristics
that can be used to suppress it: i) it is of $\nu_e$ flavor
and ii)  directional. The required rejection rate of order
$10^{-6}-10^{-9}$ is challenging in spite of the excellent
sensitivity of the detectors  to the neutrino direction,
especially for the $hep$ and $^7B$ fluxes. Flavor
identification at this level seems difficult since
$\nu_{\mu}$ or $\nu_{\tau}$ are not detected. Finally, flavor
identification is only valid in the absence of solar flavor
oscillations, an issue which is unsettled.

Above the solar cutoff the background neutrinos fluxes are
dominated by the decays of hadrons produced by interactions
of cosmic rays with atmospheric nuclei. The main contribution
is from pions with $\nu_e :\nu_{\mu} \simeq 1:2$. The
atmospheric neutrino fluxes are understood with roughly
$10\%$ precision. At the lower energies they are  strongly
dependent on the magnetic latitude and arrival directions
because of the geomagnetic cutoff for cosmic rays. Locations
with low geomagnetic latitude have  smaller atmospheric
neutrino fluxes at low energies which is an advantage if
a bound on PBHs is to be extracted.

The atmospheric flux of $\nu_{\tau}$ is multiply suppressed.
The open decay channels involve heavy parents such as $D_s$
or bottom hadrons~\cite{ds} which have smaller production
cross section than pions and kaons and branching ratios for
$\nu_{\tau}\tau$ decays which are over an order of magnitude
($4\%$ for $D_s$) below those for pions and kaons. Production
of heavy hadrons has a higher threshold energy. Because of the
steepness of the cosmic ray parent flux, increased thresholds
result in a drastic suppression of the flux at low energies.
The calculation is very similar to that of the prompt muon
neutrino flux of charm origin~\cite{vol1} and subject to
similarly large uncertainties associated with our poor
understanding of the production cross section for $D_s$ in
the region near threshold. There is no point then in
performing a detailed calculation, and we obtain a simple
estimate of the $\nu_{\tau}$ flux by rescaling the prompt
muon flux by a factor. Considering that $D_s$ production is
suppressed and that the branching ratios of the decays
$D_s \rightarrow \nu _\tau + ...$ are
smaller than the corresponding ones for the decays of other
$Ds$, we approximate the above factor to be of order 0.1.
This agrees with the estimates made for LHC~\cite{lhc}.

In Fig.~3 we compare PBH fluxes  to the
atmospheric $\nu _e$ and $\nu _\mu$ fluxes at a north U.S.
location~\cite{gaisser}, to the solar neutrino $hep$ flux
and  to our estimate of the atmospheric $\nu_{\tau}$ flux.
The PBH fluxes are calculated for $\beta =3$, because one
can assume that at the formation of PBHs the Universe is
radiation dominated. The figure allows one to identify  the
most promising regions for obtaining an improved bound. The
high energy region is promising since the background falls
faster than the signal. Unfortunately prompt neutrinos from
charm decay are expected to dominate at the highest energies
over the fluxes of pion and kaon decay origin and may thus
prevent the atmospheric neutrino flux to fall below the PBH
flux. On a more speculative note, the high energy neutrino
flux may be dominated by point sources which is not a problem
unless a large number conspire to produce a diffuse flux.
Searching for a $\nu_{\tau}$ flux seems to be most promising.
Even nonobservation is of interest as it may yield important
bounds on PBH abundances.

\section{Final stage evolution: detection of $\nu$'s from
individual black holes}

As we have seen in Sect. 2 the final stage of a PBH is
explosive. In  order to determine the present density of
these exploding black holes one must look for individual
sources. In this section we show that high energy neutrino
telescopes can significantly contribute to this search.
Their threshold energy is of order $E\geq 10~GeV$ for
$\nu_\mu$ neutrinos. The main contribution to the flux
$f(x,M)$ comes, in the case of explosions, from the decay
of fragmentation pions. This is because other contributions
are either smaller or peak at lower energies.

Because we are going to consider PBHs whose lifetimes are
short compared to a typical observation time (of order of
years), the relevant quantity is the time integrated flux
\begin{eqnarray}
{{\rm d}N_\nu \over {\rm d} E }
	&=&\int_{t_i}^{t_f} {\rm d} t f(x,M) \nonumber   \\
        & &                               \label{eq:liftfl} \\
	&=& 1.19~ 10^{39} GeV^{-1} \left({1GeV\over E}\right)^3
\int_0^{E/ kT_i} {\rm d} x x^2 \alpha ^{-1}f(x, M) , \nonumber
\end{eqnarray}
where $x=E/kT$. Here we have used Eqs.~(\ref{eq:temp}) and
(\ref{eq:massloss}) to convert the time integral into an
integral over the inverse temperature. We can neglect redshift
effects because we will find that one does not expect to see
PBH explosions from further than a few parsecs.

The $\nu _\mu$'s  interact with the Earth and the secondary
upcoming muons produce the experimental signature in the
underground detector. Only muons entering
the detector from below  are detected, because
the background of downward atmospheric muons dominates
any neutrino-induced signal. The probability  that
a neutrino produces a muon  above the threshold
energy  $E_{\scriptscriptstyle D}$ at the
detector~\cite{prob} is
\begin{equation}
P_{\nu \rightarrow \mu}(E_\nu, E_{\scriptscriptstyle D})
= N_{\scriptscriptstyle A}\int_{E_{\scriptscriptstyle D}}^{E_\nu}
dE_\mu {{\rm d}\sigma \over {\rm d} E_\mu} (E_\mu, E_\nu)
R_{\mu}(E_\mu, E_{\scriptscriptstyle D}),         \label{eq:peff}
\end{equation}
where $N_{\scriptscriptstyle A}$ is Avogadro's number,
${{\rm d}\sigma / {\rm d} E_\mu }$ is the
differential charged current cross
section~\cite{charged,totcross} and $R_\mu$ is the muon
range~\cite{range}. We updated  ${{\rm d}\sigma
/ {\rm d} E_\mu }$ of~\cite{charged,totcross} by using the
latest small-$x$ parton distribution functions of \cite{smallx}.
The evaluation of Eq. (\ref{eq:peff}) was carried out using the
Monte Carlo integration routine VEGAS~\cite{vegas}.

The predictions relevant to experiment are obtained by
integrating the muon flux in the detector above
$E_{\scriptscriptstyle D}$. The number of observed events
is given by:
\begin{equation}
N_\nu (E_{\scriptscriptstyle D}, \theta ) = {A\over 4\pi r^2}
\int_{E_{\scriptscriptstyle D}}
^\infty {\rm d} E ~P_{\nu \rightarrow \mu}\exp \left[
-\sigma_{\rm tot}(E_\nu)N_{\scriptscriptstyle A} X(\theta )\right]
{{\rm d}N_\nu \over {\rm d}E},       \label{eq:neno}
\end{equation}
where $r$ is the distance to the PBH, $A$ is the sensitive
area of the detector, the exponential factor accounts for
absorption of neutrinos along the chord of the Earth
$X(\theta )$ and $\sigma _{\rm tot }$ is the total cross
section~\cite{totcross}. Here $\theta $ is the zenith angle
($0\leq \theta \leq 90^{\circ }$). The dominant contribution
to the integral (\ref{eq:neno}) comes from and is evaluated
for nearly vertical neutrinos ($\theta \simeq 0$), because
only neutrinos with small $\theta $ travel far enough in
the Earth to interact and produce a muon. Detectors under
construction may, eventually, reach an effective area of
order $A\simeq 1~{\rm km}^2$. The threshold energy can be
anywhere between a few GeV to a few TeV.

Because the number of events is inversely proportional to $r^2$,
there is a maximum $r$ for which the PBH is detectable. This
distance depends on the initial temperature of the PBH and is
increased for lower temperatures because of the increased
observation time.  Fig.~4  shows this maximum distance as a
function of the detector threshold energy
$E_{\scriptscriptstyle D}$ for  initial temperatures ranging
from 50~GeV to 20~TeV. The limiting distance is determined by
the requirement that we detect at least 10 muons per burst.

For a PBH to be detectable it has to stand out over background.
In this case the high energy atmospheric flux  is the relevant
background. Our calculation follows Ref.~\cite{volkova}, and is
in agreement with other calculations and measurements of the
atmospheric neutrino background~\cite{gaisser}. The number of
background events in the detector is:
\begin{equation}
N_{\rm bgr}=\tau ~A \int_{E_{\scriptscriptstyle D}}^{\infty}
{\rm d}E~\omega _{\scriptscriptstyle D}(E)~
{{\rm  d}^4~N
\over {\rm d}A~{\rm d}t~{\rm d}\omega ~{\rm d} E}~P_{\nu
\rightarrow  \mu} \ ,
					\label{eq:back}
\end{equation}
where $\omega _{\scriptscriptstyle D}$ is the angular
resolution of the detector
\begin{equation}
\omega _{\scriptscriptstyle D} (E) = {0.314~{\rm sr}
\over E({\rm GeV})} \ ,
\end{equation}
and $\tau$ is the lifetime of the black hole. For detectability
we require that the signal (\ref{eq:neno}) exceeds the
background (\ref{eq:back}) by  $ 5\sigma$. The maximum
distance for which this condition holds, is shown on Fig.~4
as a function of the detector energy for initial temperatures
ranging from 50~GeV to 20~TeV.

The distance of an observable PBH must fall below both curves
on Fig.~4.  For $kT_i\geq 1~{\rm TeV}$ the 10 muon signal
condition is stronger. This is because the lifetime of such a
PBH is so short that few background events can accumulate
during its observation. For lower initial temperatures the
10 muon signal condition is stronger for high detector thresholds,
while the $5\sigma$ condition is stronger for low thresholds.
This combined condition is shown on Fig.~5. The optimal condition
is obtained for a PBH with $kT_i\simeq 100~{\rm GeV}$. The
lifetime of such a PBH is about 1.5 months, longer than the
lifetime of PBHs usually referred as ``explosive'' in the
literature, but still small compared to the detection time.
The maximum distance is $2.3\times 10^{-3}{\rm pc}$ at
$E_{\scriptscriptstyle D}\simeq 140~{\rm GeV}$,
which gives an upper bound
$1.2\times 10^7 {\rm pc}^{-3} {\rm yr}^{-1}$ on the number
of PBHs exploding in our neighborhood per year. Here we
used detector solid acceptance angle $2\pi$, observational
time 3.3 years, and assumed a uniform PBH distribution in
our neighborhood. The detector acceptance solid angle is
$2\pi$ because only upward muons have experimental signature.
In order to compare our result to existing experimental
bounds, we choose the observation time to be that of the
search establishing the strongest bound.  It is
$8.5\times 10^5 {\rm pc }^{-3}{\rm yr}^{-1}$ from high
energy $\gamma $ ray measurement from $kT_i \simeq 10$~TeV
PBH, and it is of two order of magnitude smaller than that
of any other direct search~\cite{cygnus}.

Our result is stronger than any existing experimental
bounds, except that of \cite{cygnus}. Our bound is
about an order of magnitude larger, but can be interesting
if for some reason detection of neutrinos is preferable
over detection of $\gamma $ rays from PBHs. The issue has
indeed been raised whether $\gamma $ rays are emitted
as independent particles in the presence of thermal
electrons surrounding the black hole~\cite{heckler}.

\section{Detector considerations}

It  obviously remains to be seen whether detectors will ever
reach  the sensitivities to observe the neutrino fluxes
predicted in our calculations. In the diffuse case, the
challenge is formidable as can be seen from Fig.~3. One must
improve on the sensitivity to detect atmospheric neutrinos by
over three orders of magnitude even when restricting
observations to the favorable vertical neutrinos. The most
promising search involves tau neutrinos with energy below
$10$~GeV. Detecting the $\tau$-lepton emerging from a charged
current interaction may be possible since the neutrino energy
can exceed the $\tau$-lepton mass but this will certainly
represent a challenge. Alternatively one may look for the
difference between the total neutrino flux and those of
electron and muon flavor. This procedure requires a precision
of $0.1\%$ which is probably out of question in the near
future. The search for explosions presents us, in the end,
with the most realistic opportunities.

\section{Conclusion}

In this paper we have calculated the neutrino fluxes from
PBHs and their detectability for both low and high energies.
We have updated the previous calculations by including
contributions not only from direct neutrino emission, but
from muon and tau decays and from quark fragmentation. The
latter is especially important because of the large
number of quark degrees of freedom.

In the diffuse case the  fluxes are
increasing functions of $\beta $ in the allowed range
$2\leq \beta  \leq 3$. The $\nu _\mu$ and $\nu _e$
fluxes are dominated by the fragmentation contribution
for $E_\nu$ less than about 100~MeV. Around 1~MeV it
is about an order of magnitude higher than the other
contributions. For larger energies all contributions
decrease as $E_\nu ^{-3}$, and the total flux is dominated
by the direct emission of neutrinos. In the case of
$\nu _\tau$ the direct flux dominates at all energies,
because there is no relevant fragmentation contribution.
In the individual PBH case the time integrated flux is
relevant. The $\nu _\mu $ contribution from the first
decay of fragmentation pions is the largest.

We have compared the calculated fluxes to the relevant
background to determine which neutrino species at
what energy range has large enough flux to be detectable.
In the diffuse case the $\nu_e$ and $\nu _\mu$ fluxes are
larger than the atmospheric background only for $E_\nu$
smaller than a few MeV. For $\nu _e$ the huge directional
solar $hep$ background can cause further observational
difficulties. Most promising is the $\nu _\tau$ case,
because the only relevant background, the atmospheric
$\nu _\tau$ flux, is highly suppressed.

Our best chance is to look for individual PBHs in our
neighborhood. We have shown that the new kilometer-scale,
high energy ($E_{\scriptscriptstyle D}\simeq 1~{\rm GeV}
-1~{\rm TeV}$) neutrino telescopes can play a significant
role in this search. They are sensitive to $\nu _\mu$'s.
Furthermore at these high energies the only background,
the atmospheric neutrino background, is  suppressed.
The bound that can be established with these detectors is
comparable to existing experimental bounds, obtained by
searching for high energy $\gamma $ ray bursts.

\section*{Acknowledgements}

We thank Jesus Armada for his contribution to the early
stages of this work, Ricardo Vazquez and Brian Wright for
interesting discussions and Brian Wright for careful
reading of the manuscript. This work was supported in
part by the University of Wisconsin Research Committee
with funds granted by the Wisconsin  Alumni Research
Foundation, in part by the U.S.~Department of Energy under
Contract No.~DE-AC02-76ER00881, in part by the Texas
National Research Laboratory Commission under Grant
No.~RGFY9173 and in part by the Xunta de Galicia,
reseach contract XUGA 20604A93.

\newpage

\section*{Figure captions}
\vskip 1cm
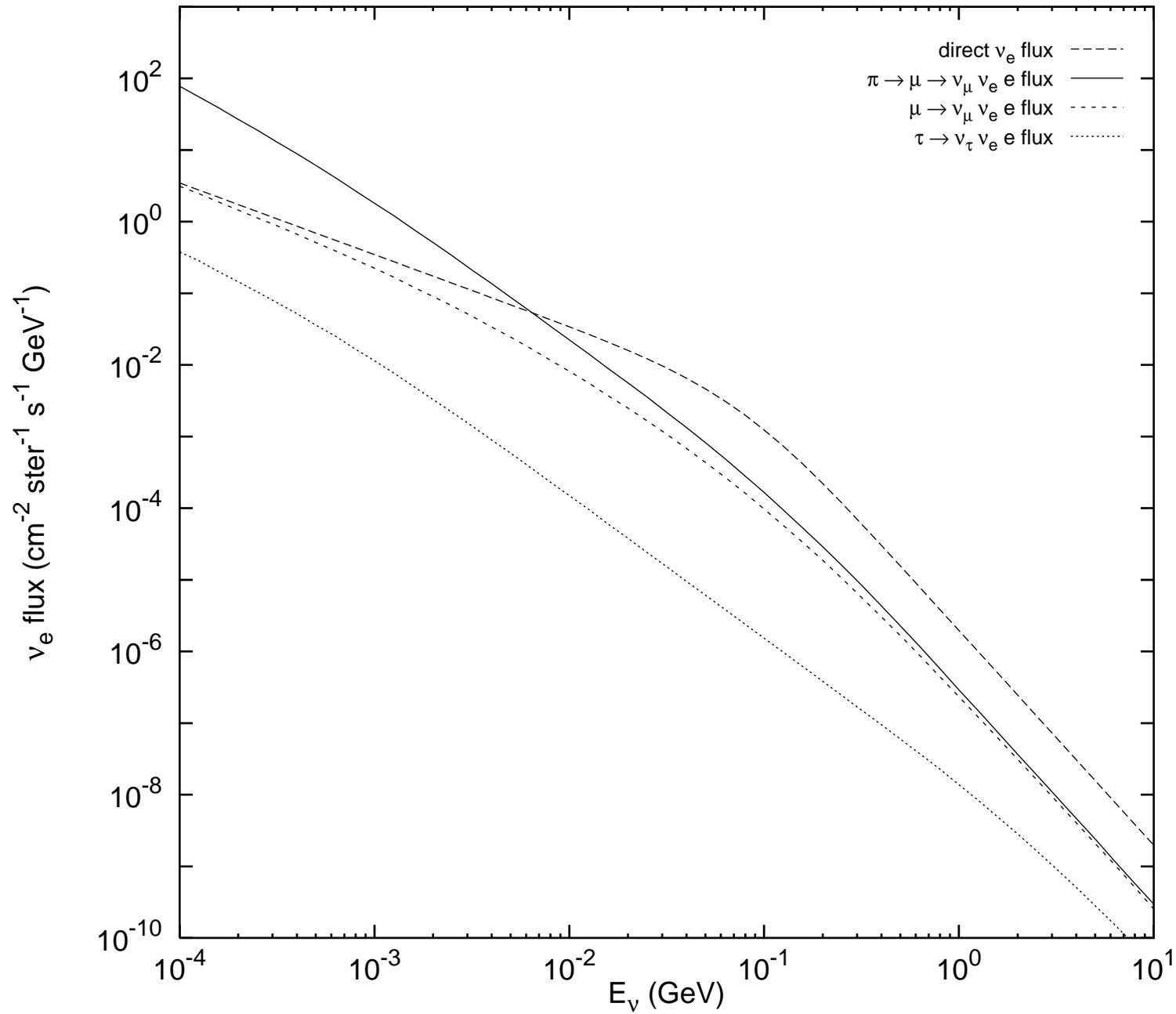
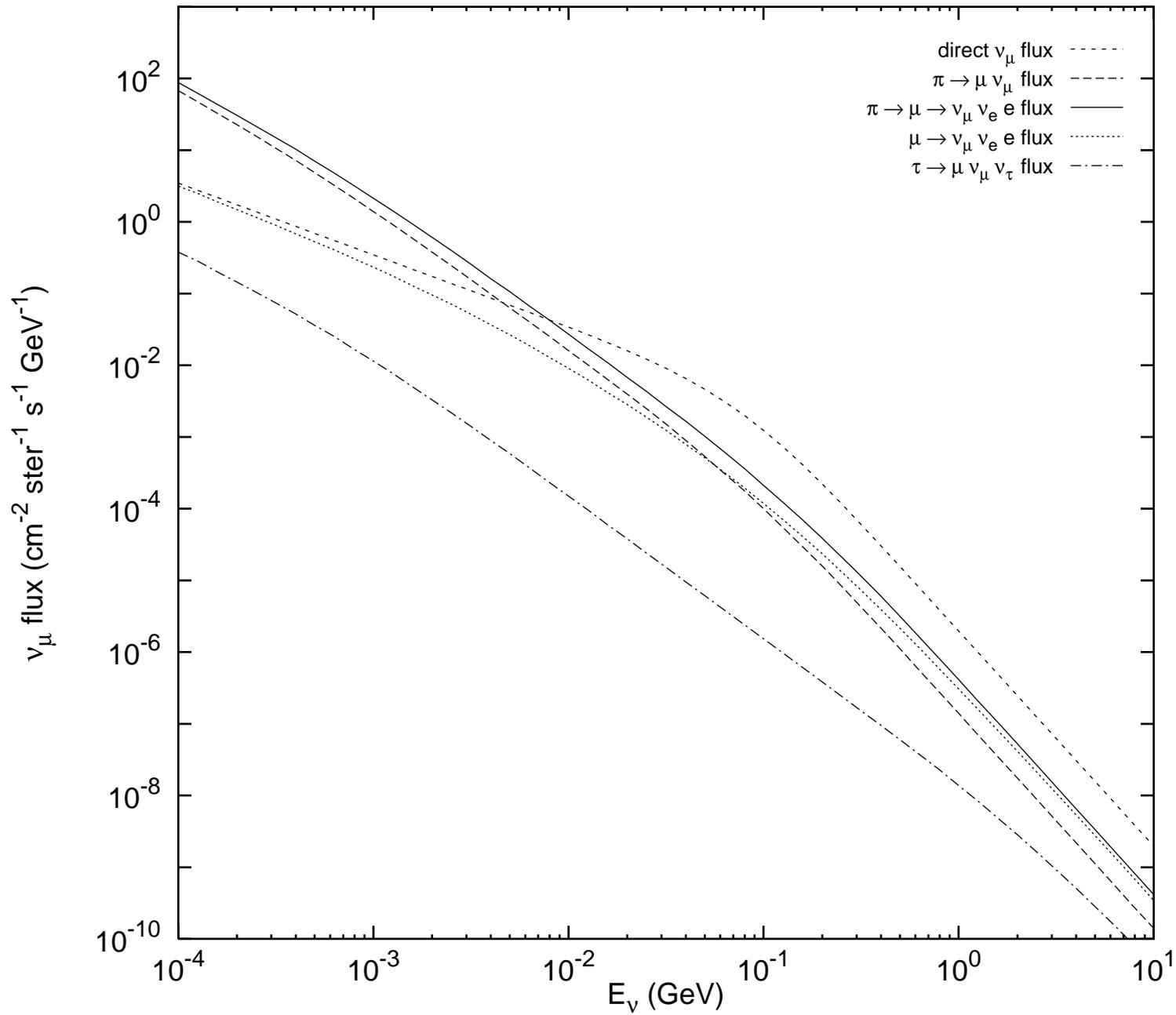
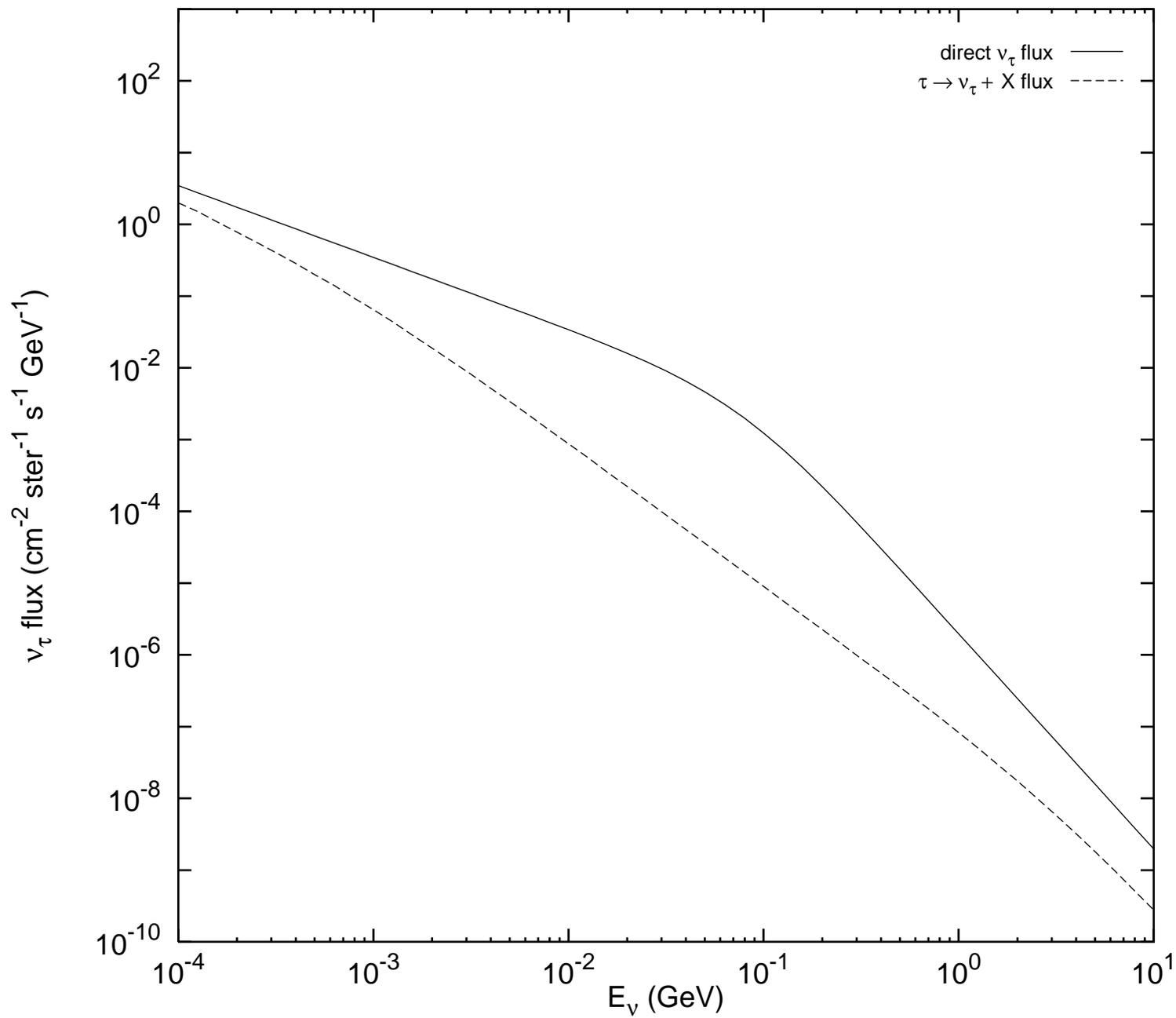
\noindent {\bf Fig. 1.}  Contributions to diffuse neutrino flux
(a) for $\nu_e$, (b) for $\nu _\mu $ and (c) for $\nu _\tau $. \\
\vskip .5cm
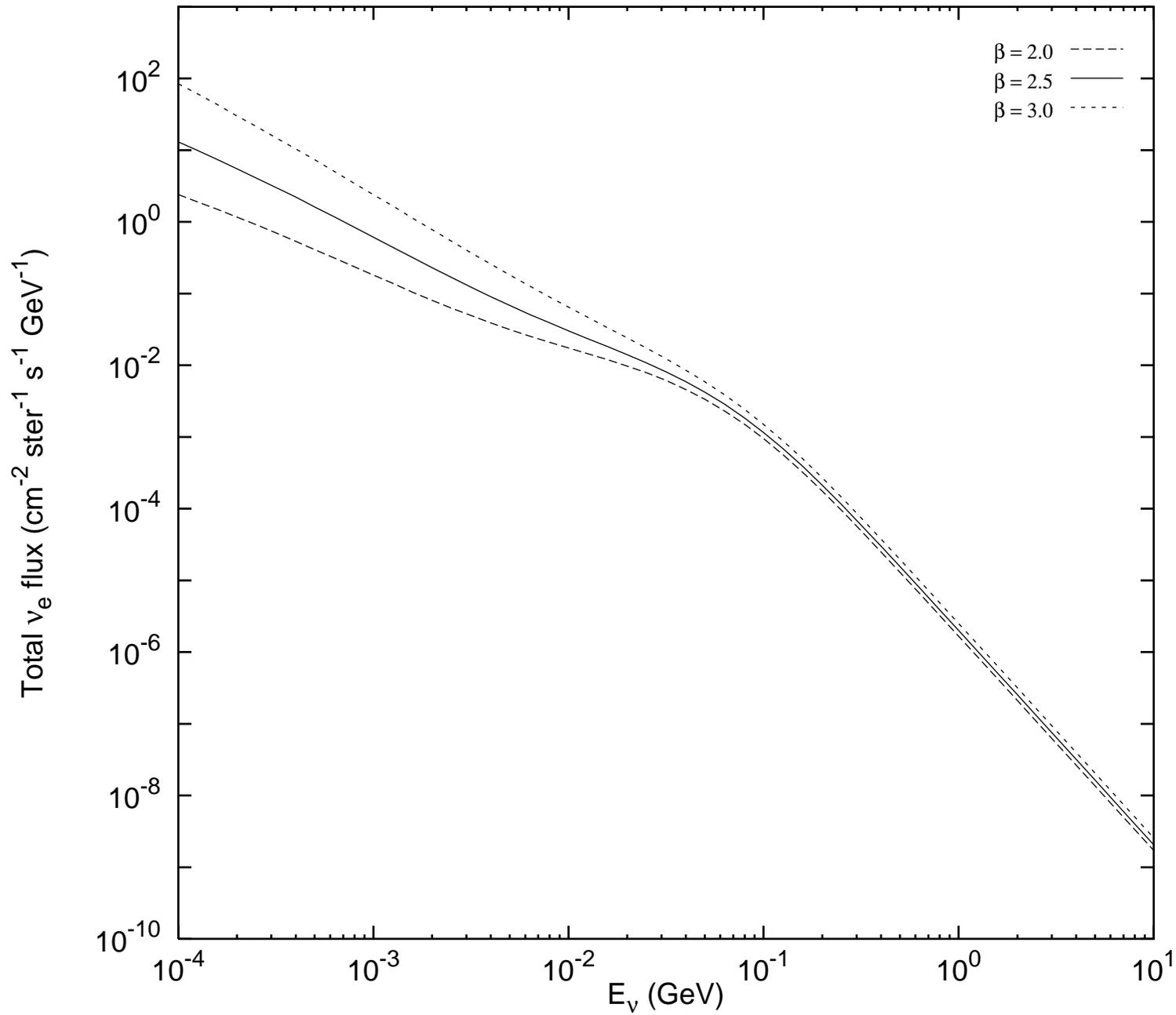
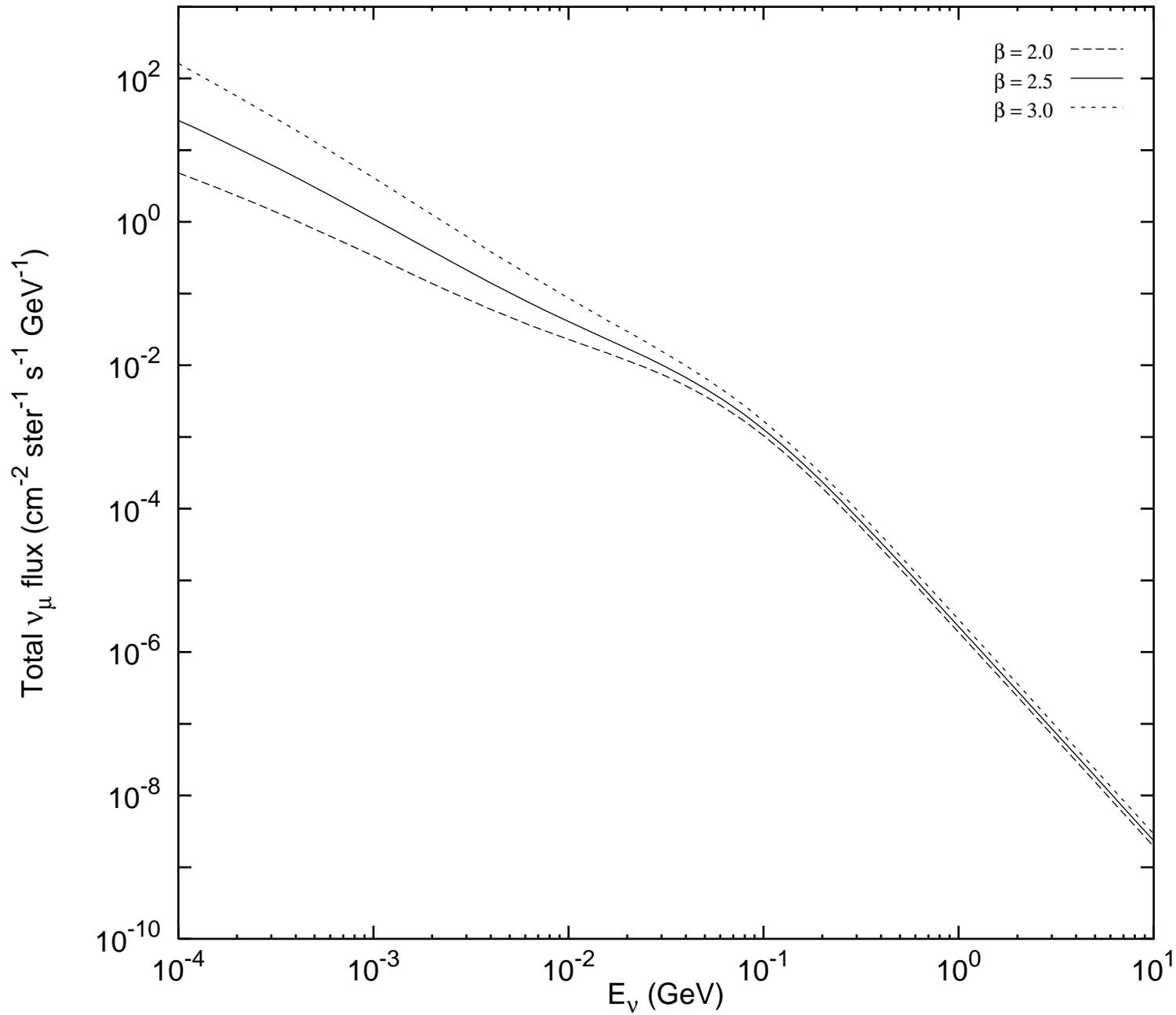
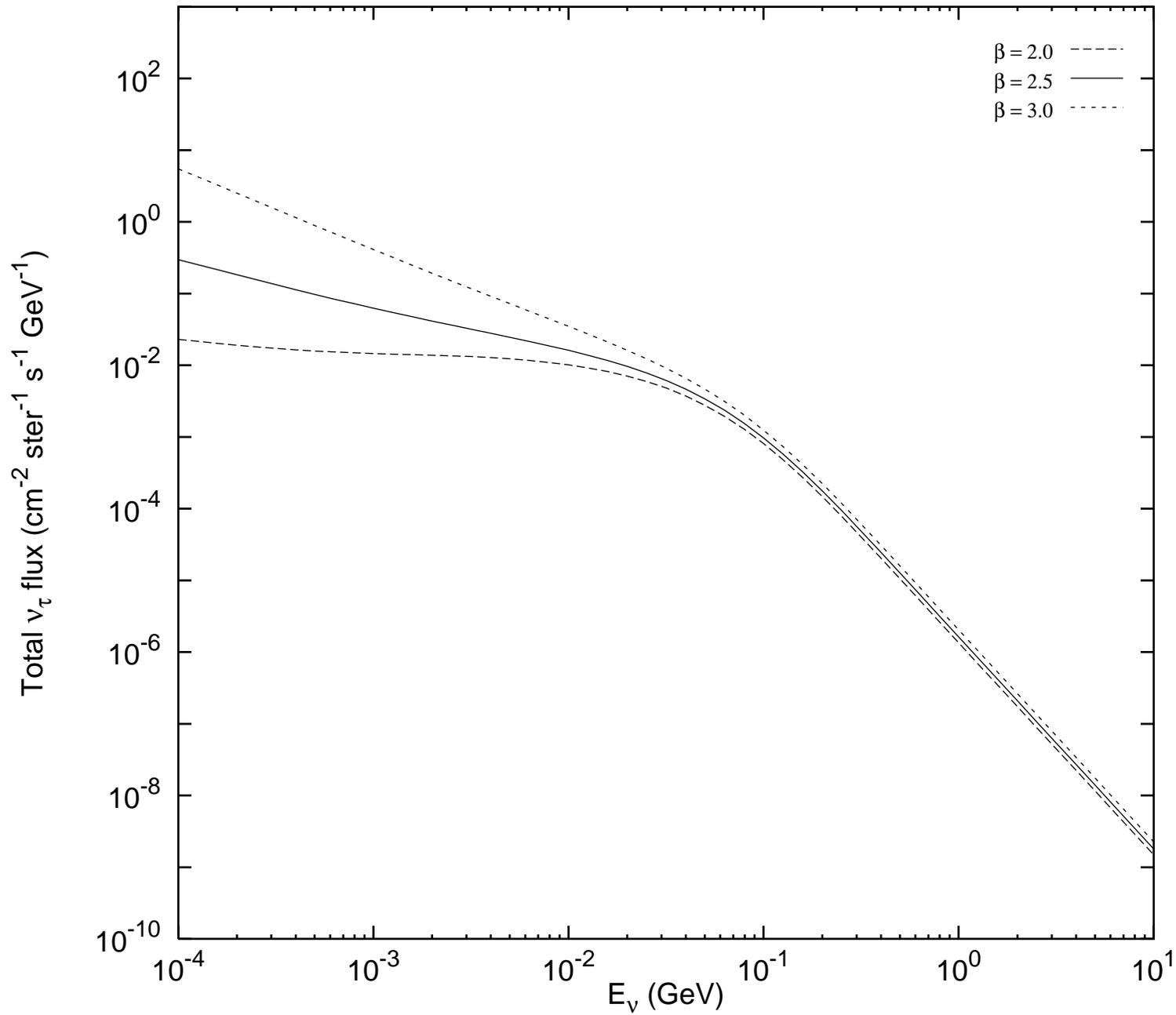
\noindent {\bf Fig. 2.} Total diffuse neutrino fluxes for
$\beta $= 2, 2.5 and 3 and (a) for $\nu_e$, (b) for $\nu _\mu $
and (c) for $\nu _\tau $.  \\
\vskip .5cm
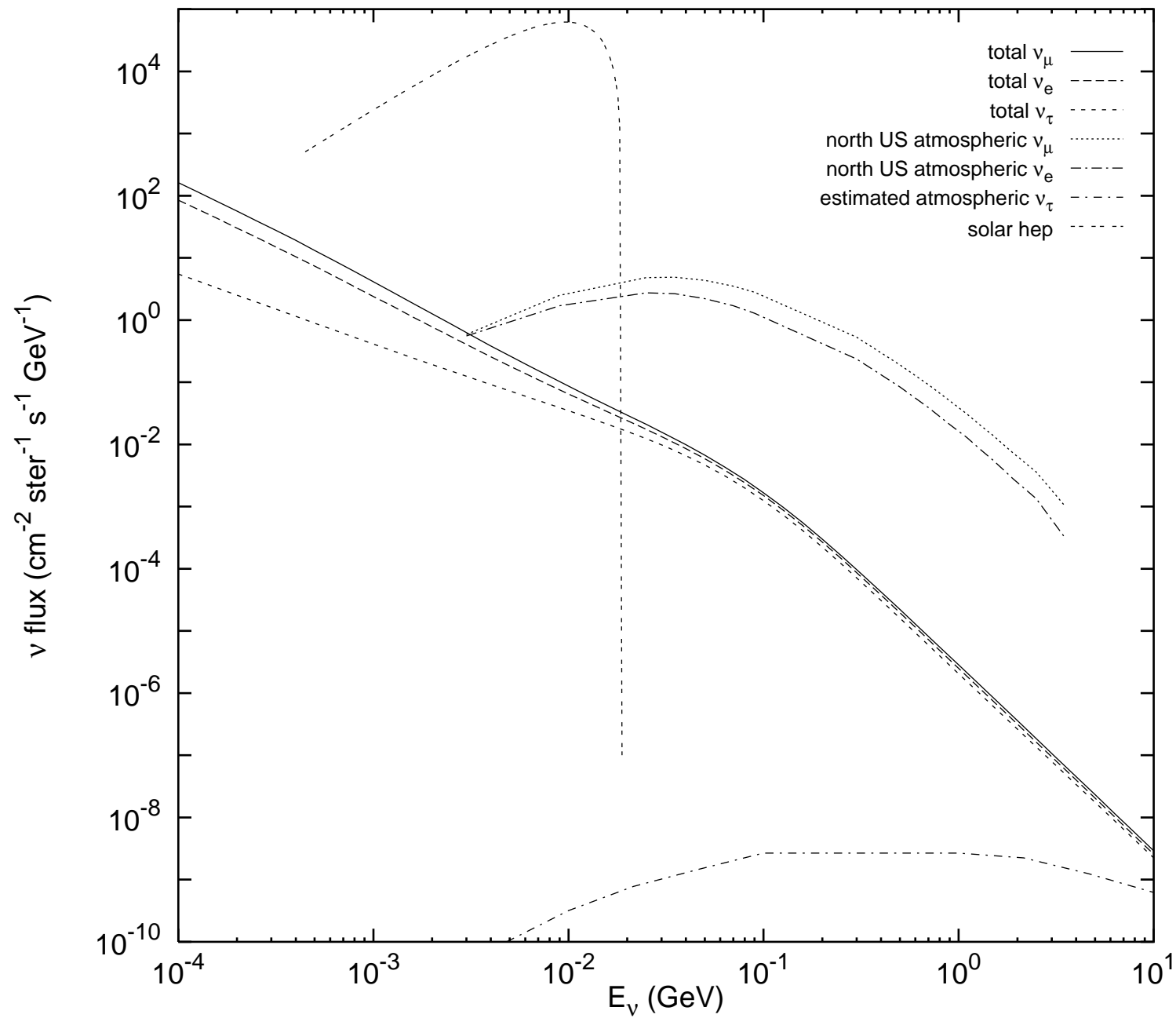
\noindent {\bf Fig. 3.}  Total diffuse neutrino fluxes for
$\beta $=3 compared to the atmospheric neutrino backgrounds
and to the solar $hep$ flux.  \\
\vskip .5cm
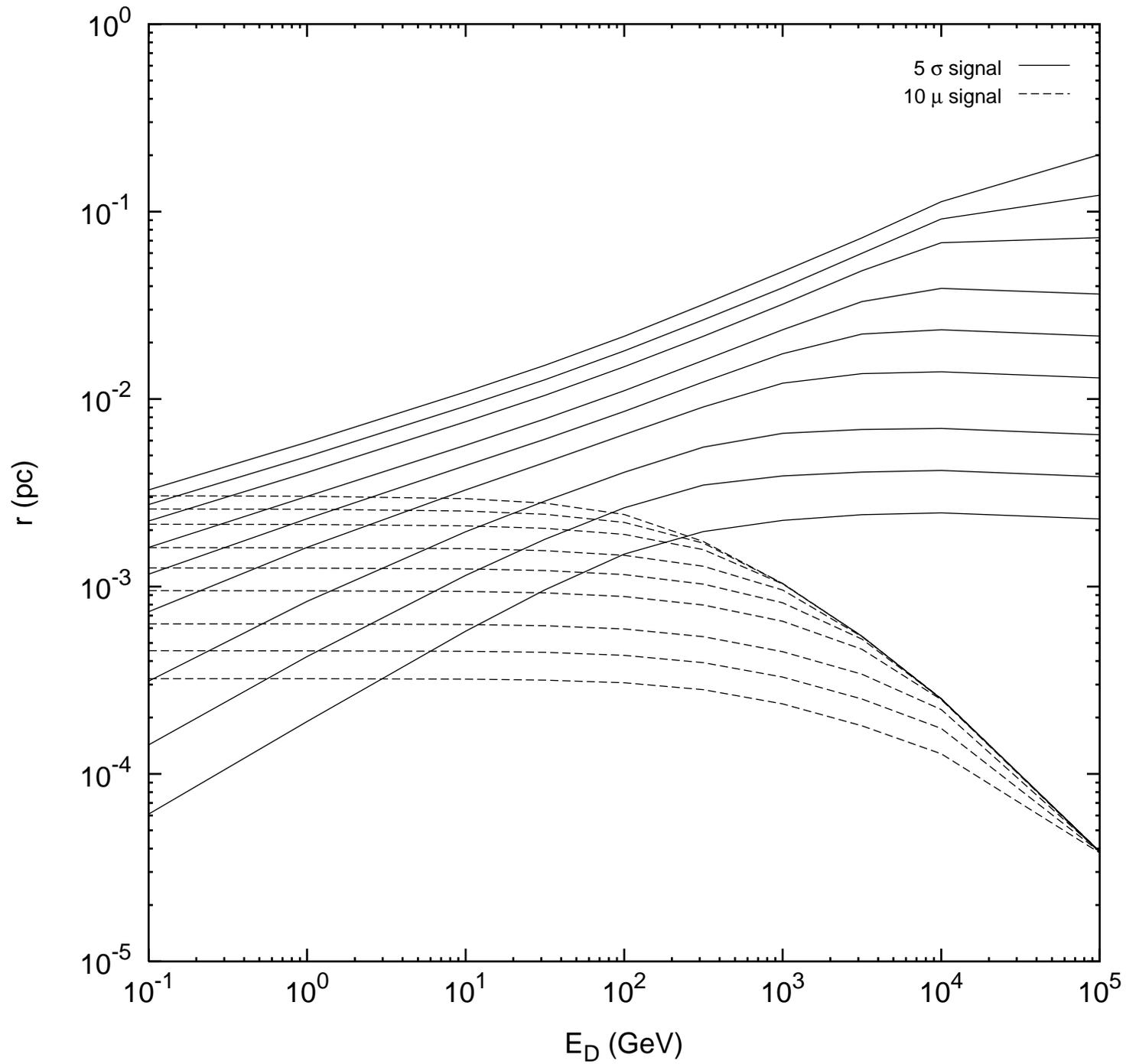
\noindent {\bf Fig. 4.} The 10 muon signal (dotted lines) and
5$\sigma$ signal (solid lines) conditions as the function of the
detector threshold energy, $E_{\scriptscriptstyle D}$.
Dotted lines correspond to $kT_i = $ 50~GeV, 100~GeV, 200~GeV,
500~GeV, 1~TeV, 2~TeV, 5~TeV, 10~TeV and 20~TeV from top to
bottom, solid lines correspond to the same $kT_i$ values from
bottom to top. \\
\vskip .5cm
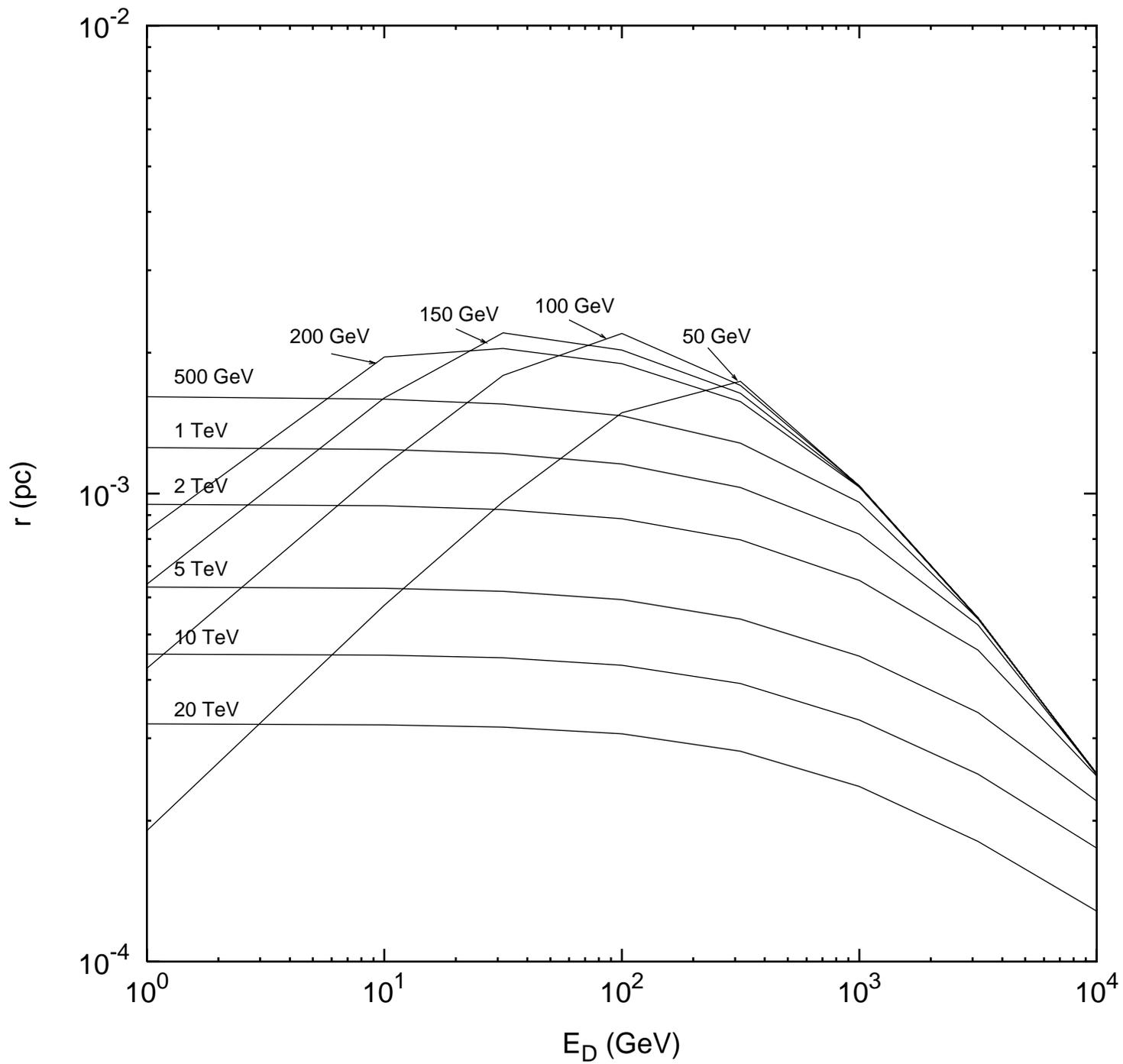
\noindent {\bf Fig. 5.} Combined 10 muon signal and 5$\sigma $
signal conditions as the function of the detector threshold
energy $E_{\scriptscriptstyle D}$ for $kT_i = $ 50~GeV, 100~GeV,
150~GeV, 200~GeV, 500~GeV, 1~TeV, 2~TeV, 5~TeV, 10~TeV
and 20~TeV.  \\
\end{document}